\documentclass[aps,amsmath,amssymb,prl,twocolumn,letterpaper,showpacs]{revtex4-1}
\usepackage{graphicx}
\usepackage{bm}
\usepackage{txfonts}
\usepackage{mathrsfs}
\usepackage{feynmf}
\usepackage{comment}
\usepackage{epsfig}

\newcommand{\al}{\alpha}
\newcommand{\be}{\beta}
\newcommand{\g}{\gamma}
\newcommand{\de}{\delta}

\newcommand{\z}{\zeta}

\newcommand{\thi}{\theta}

\newcommand{\ka}{\kappa}
\newcommand{\la}{\lambda}

\newcommand{\p}{\pi}

\newcommand{\s}{\sigma}

\newcommand{\w}{\omega}
\newcommand{\W}{\Omega}
\newcommand{\De}{\Delta}

\newcommand{\pd}{\partial}

\newcommand{\round}[1]{\left( #1 \right)}
\renewcommand{\square}[1]{\left[ #1 \right]}

\newcommand{\beq}{\begin{equation}}
\newcommand{\eeq}{\end{equation}}
\newcommand{\Beq}{\begin{eqnarray}}
\newcommand{\Eeq}{\end{eqnarray}}
\newcommand{\bml}{\begin{multline}}

\newcommand{\bea}{\begin{align}}
\newcommand{\ena}{\end{align}}
\newcommand{\bsp}{\begin{split}}
\newcommand{\esp}{\end{split}}
\newcommand{\down}{\downarrow}
\newcommand{\up}{\uparrow}

\newcommand{\bS}{{\boldsymbol{S}}}

\newcommand{\ex}{\hat{\boldsymbol x}}
\newcommand{\ey}{\hat{\boldsymbol y}}
\newcommand{\ez}{\hat{\boldsymbol z}}

\newcommand{\bj}{{\boldsymbol j}}

\newcommand{\bn}{{\boldsymbol n}}

\newcommand{\bE}{{\boldsymbol E}}
\newcommand{\sE}{\mathcal{E}}
\newcommand{\bve}{{\boldsymbol \varepsilon}}

\newcommand{\bx}{\boldsymbol{x}}

\newcommand{\bs}{\boldsymbol{s}}

\newcommand{\bta}{\boldsymbol{\tau}}

\newcommand{\tg}{\tilde{\gamma}}

\newcommand{\tia}{\tilde{a}}

\newcommand{\sA}{\mathscr{A}}

\begin{document}
\title{Nonlocal AC Magnetoresistance Mediated by Coherent Spin Excitations}
\author{So Takei}
\affiliation{Department of Physics and Astronomy, University of California, Los Angeles, California 90095, USA}
\author{Yaroslav Tserkovnyak}
\affiliation{Department of Physics and Astronomy, University of California, Los Angeles, California 90095, USA}
\date{\today}

\begin{abstract}
Electrical response of two diffusive metals is studied when they are linked by a magnetic insulator hosting topologically stable (superfluid) spin current. We discuss how charge currents in the metals induce a spin supercurrent state, which in turn generates a magnetoresistance that depends on the topology of the electrical circuit. This magnetoresistance relies on phase coherence over the entire magnet and gives a direct evidence for spin superfluidity. We show that driving the magnet with an ac current allows coherent spin transport even in the presence of U(1)-breaking magnetic anisotropy that can preclude dc superfluid transport. Spin transmission in the ac regime shows a series of resonance peaks as a function of frequency. The peak locations, heights and widths can be used to extract static interfacial properties, e.g., the spin-mixing conductance and effective spin Hall angle, and to probe dynamic properties such as the spin-wave dispersion. Ac transport may provide a simpler route to realizing nonequilbrium coherent spin transport and a useful way to characterize the magnetic system, serving as a precursor to the realization of dc superfluid spin transport.
\end{abstract}
\maketitle

%%%%%%%%%%%%%%%%%%%%%%%%%%%%%%%%%%%%%%%%%%%%%%%%%%%%%%%%%%%
{\em Introduction}.|Understanding spin transport via collective magnetic excitations is currently gaining attention~\cite{kajiwaraNAT10,*bauerPHYS11,*zhangPRL12,*hahnEPL14,*wangPRL14,*moriyamaAPL15,*takeiCM15,*cornelissenCM15}. An exciting frontier explores how analogs of conventional superfluidity, as observed in liquid $^4$He, can be obtained in magnetic systems, and how dissipationless spin currents can be realized and detected in such systems~\cite{soninJETP78,*soninAP10}. Conventional superfluidity is characterized by a rigid U(1) order parameter, a single quantum-mechanical wave function describing a macroscopic number of constituent particles. Dissipationless current, being proportional to the gradient of the U(1) phase, appears only in the phase-coherent state with broken gauge invariance. A certain class of magnetic insulators, such as ferromagnetic insulators with easy-plane magnetic anisotropy, are also characterized by a U(1) order parameter, whose magnitude and phase characterize the magnetic order within the easy plane. Here, dissipationless spin current (polarized out of plane) is triggered by a collective reorientation, i.e., a planar spiraling texture of the magnetic order. Theoretical proposals for realizing and detecting such superfluid spin transport have been put forth for (ferro- and antiferro-) magnetic insulators~\cite{takeiPRL14,takeiPRB14,chenPRB14b} and multiferroic materials~\cite{chenPRB14,*chenPRL15}. %

Superfluid spin transport may be detected using a two-terminal spin transport setup, in which two spin-orbit coupled metals are attached on opposite ends of the magnetic insulator and direct (inverse) spin Hall effect facilitates spin injection (detection)~\cite{takeiPRL14,takeiPRB14}. While the two-terminal spin conductance so obtained can reveal a smoking-gun signature of spin superfluidity|with Gilbert damping spin supercurrent decays algebraically over space while spin current carried by incoherent thermal magnons decays exponentially over the spin diffusion length|such probe requires a study of spin transmission through various sample sizes. Additional signatures of superfluid transport obtainable from a single sample is desirable. On another note, a source of difficulty in realizing dc superfluid spin transport are magnetic anisotropies that break U(1) symmetry crucial for genuine superfluidity. Superfluid spin transport in the dc regime requires the U(1) magnetic order parameter to make full $2\p$ rotations within the U(1) plane. This dynamics can be quenched by such anisotropies which tend to pin the texture. While this pinning may be overcome by injecting large enough spin currents~\cite{soninJETP78}, achieving such large spin currents does pose experimental challenges~\cite{takeiPRB14}. 

In this Letter, we theoretically propose measurements based on the two-terminal setup that can help overcome these challenges and thus expedite the realization and verification of superfluid spin transport. First, we show that a nonequilibrium spin superfluid state induces unique nonlocal magnetoresistance signatures in the metallic contacts used for spin injection and detection. Due to phase coherence over the entire magnet, the nonequilibrium superfluid state is self-consistently determined by boundary conditions defined at the injection and detection interfaces. In the presence of magnetoelectric coupling at these interfaces, the charge currents in the metallic contacts define a particular superfluid state, which in turn governs a nonlocal magnetoresistance in the circuit. This phenomenon, which we refer to as spin-superfluid magnetoresistance (SSMR), depends intimately on the topology of the external electrical circuit connecting these contacts. The observation of SSMR would constitute a strong signature of spin superfluidity even without sample-size dependence studies. 

In the latter half of the work, we consider spin transport by coherent magnetic textures in the ac regime and in the presence of U(1)-breaking inplane magnetic anisotropy. Even if the dc superfluid transport is quenched by pinning, coherent ac spin transport can still occur via oscillations of the local magnetic order about the pinning potential minimum (without the need for full $2\p$ rotations as in the dc regime), thus circumventing the pinning problem. Spin transmission through a magnet of length $L$ shows resonances as a function of ac frequency $\w$. For $\w\gg\w_0$, $\hbar\w_0$ being the gap in the spin-wave spectrum associated with the inplane anisotropy, the resonance peaks occur in intervals of $\p v/L$, allowing one to extract the spin-wave velocity $v$. Spin transmission is exponentially suppressed for $\w\ll\w_0$, and the position of the first peak (at $\w=\w_0$) gives a direct measure of the gap. Furthermore, measurements of peak heights and widths allow one to extract the effective spin Hall angle and spin-mixing conductance at the interfaces. We show that finite-frequency spin waves can support spin transmission decaying algebraically as a function of sample length, as in spin transmission via dc spin superfluidity. Therefore, the ac transport studies should not only serve as a simpler route to realizing nonequilbrium coherent spin transport, but also as a useful way to characterize the magnetic system and a meaningful precursor to the ultimate realization of the dc superfluid spin transport.

%The ac frequency brings a new energy scale to the problem, and can be used to extract spectral properties of the bulk spin waves, e.g., the spectral gap governed by the U(1)-breaking anisotropies. We show that the ac spin transport can still be used to quantify static interfacial properties such as the effective spin-mixing conductance and spin Hall angle at the metal$|$magnet interfaces, as well as the Gilbert damping parameter in the magnet's bulk. 

\begin{figure}[t]
\centering
\includegraphics*[width=0.97\linewidth]{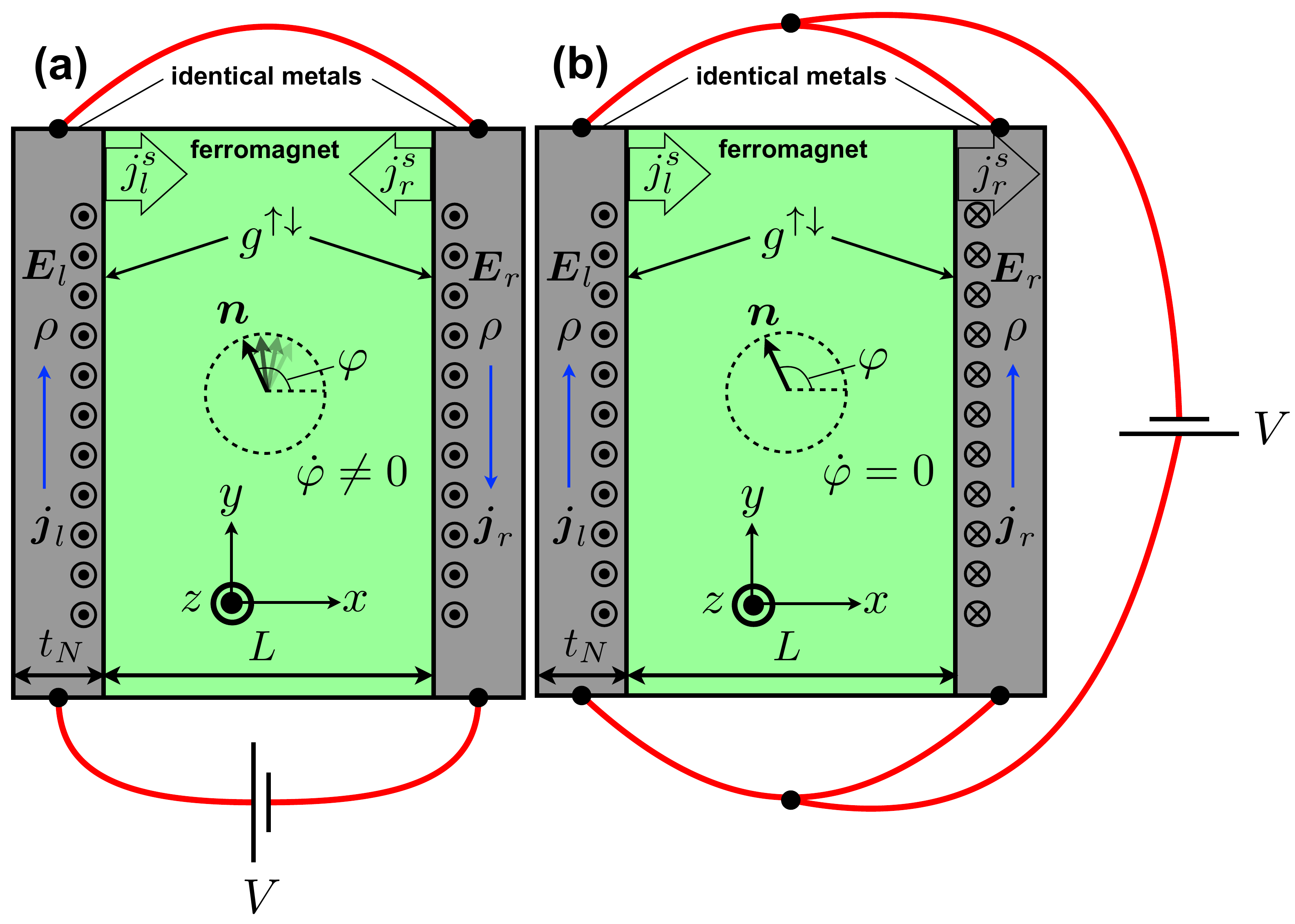}
\caption{(color online) Schematics of the series (a) and parallel (b) configurations, as detailed in the text.}
\label{config}
\end{figure}

{\em General considerations}.|Consider a magnetic insulator sandwiched by two normal metals as sketched in Fig.~\ref{config}. Our focus is on magnets well below the magnetic ordering temperature, which, in the long-wavelength limit, is characterized by a slow continuum variable, the U(1) order parameter $\bn(\bx,t)$, encoding magnetic state in the (easy) $xy$ plane. Specifically, recall that for a ferromagnetic insulator, $\bn$ corresponds to the direction of the local spin density, and the U(1) easy plane can generally be defined by the shape anisotropy~\cite{takeiPRL14}. For an isotropic antiferromagnet, $\bn$ is the direction of the local N\'eel order, and the U(1) plane is defined to be normal to a uniform external magnetic field~\cite{takeiPRB14}. For the axially symmetric magnetic state, the spin density polarized along the $z$ axis is a hydrodynamic quantity that is approximately conserved. (Its relaxation is in practice governed by spin-orbit impurities, which microscopically break the symmetry.) For simplicity, we take the normal metals and the interfaces to be identical on the two sides. The metals, treated here as diffusive films of thickness $t_N$ lying parallel to the $yz$ plane, possess strong spin-orbit coupling with an effective spin Hall angle $\thi$ at the interfaces.

The left ($l$) and right ($r$) interfaces, located at $x=0$ and $x=L$, respectively,  break translational symmetry along the $x$ direction, but full translational symmetry is assumed parallel to the interface ($yz$) plane. The entire heterostructure can thus be described using three coarse-grained hydrodynamic fields: the U(1) phase $\varphi(x,t)$ and out-of-plane spin density $s_z(x,t)$ in the magnet and the 2D charge current densities in the left and right normal metals, $\bj_l(t)\equiv(j^y_l,j^z_l)$ and $\bj_r(t)\equiv(j^y_r,j^z_r)$. For concreteness, we hereafter focus on an easy-plane ferromagnet~\cite{takeiPRL14}: $\bn\equiv\bs/s$, in this case, being the direction of the local spin density $\bs\approx(s\cos\varphi,s\sin\varphi,s_z)$, where $s$ is the magnitude of the equilibrium spin density associated with the magnetic order. The formalism is sufficiently general that it can be readily extended to other magnets supporting spin superfluidity; it is straightforward to show, in particular, that the case of a Heisenberg antiferromagnet is closely analogous \cite{takeiPRB14}. The dynamics of an isolated easy-plane ferromagnet is given by~\cite{takeiPRL14}
\beq
\label{EOMdamp}
\dot\varphi=\frac{K}{s}n_z+\al\dot n_z\ ,\quad\dot n_z=\frac{A}{s}\varphi''-\al\dot\varphi\ ,
\eeq
where $A$ and $K$ parameterize the exchange stiffness and the easy-plane magnetic anisotropy, respectively, and $\al$ is the Gilbert damping parameter.
%The order parameter is parameterized by the azimuthal angle $\varphi$ and its $z$ projection $n_z$ via $\bn=[(1-n_z^2)^{1/2}\cos\varphi,(1-n_z^2)^{1/2}\sin\varphi,n_z]$, and
The primes (dots) denote differentiation with respect to $x$ (time). Recognizing the second equation in Eq.~\eqref{EOMdamp} a the continuity equation for $s_z\equiv sn_z$, the $z$-polarized spin current (hereafter referred to as simply spin current) reads $j^s(x,t)=-A\varphi'(x,t)$. 

In the presence of an external electric field $\bE$, a uniform current-carrying state of an isolated metal is governed by Ohm's law $\rho\bj(t)=\bE(t)$, where $\rho$ is its (2D) resistivity. In the presence of spin-orbit coupling at metal$|$magnet interfaces, current in the metal can induce a torque $\bta$ on the adjacent ferromagnetic moments, and, inversely, the ferromagnetic dynamics would induce an electromotive force in the adjacent metal. According to spin Hall phenomenology~\cite{tserkovnyakPRB14}, the torques at the left and right interfaces can be written as
\beq
\label{torque}
\bta_{l,r}=\pm(\eta+\vartheta\bn_{l,r}\times)(\ex\times\bj_{l,r})\times\bn_{l,r}\ ,
\eeq
the upper (lower) sign corresponding to the left (right) interface, and constants $\eta$ and $\vartheta$ quantifying the field-like and damping-like torques, respectively. Here, $\bn_l(t)\equiv\bn(x=0,t)$ and $\bn_r(t)\equiv\bn(x=L,t)$. The coefficient for the damping-like torque can be related to the effective interfacial spin Hall angle $\theta$ via $\vartheta\equiv\hbar\tan\theta/2e t_N$~\cite{tserkovnyakPRB14}. By the Onsager reciprocity, the torque in Eq.~\eqref{torque} gives rise to an electromotive force $\bve_{l,r}$ in the adjacent metals, thereby modifying the Ohm's law to $\rho\bj_{l,r}=\bE_{l,r}+\bve_{l,r}$, where 
\beq
\label{mf}
\bve_{l,r}=\pm[(\eta+\vartheta\bn_{l,r}\times)\dot\bn_{l,r}]\times\ex\ .
\eeq
In the following, we will retain only the $y$ components of these electromotive forces, as the $z$ components are counteracted by an electrostatic buildup along the $z$ axis (supposing the magnetic dynamics are slow compared to the relevant RC time of the metallic terminals).

In addition, a physical contact to the adjacent metals gives rise to an interfacial contribution to Gilbert damping for the ferromagnet. This damping modifies the torques to $\bta_{l,r}\rightarrow\bta'_{l,r}\equiv\bta_{l,r}-\g^{\up\down}\bn_{l,r}\times\dot\bn_{l,r}$, where $\g^{\up\down}\equiv\hbar g^{\up\down}/4\p$ and $g^{\up\down}$ is the effective (interfacial) spin-mixing conductance. The torques $\bta'_{l,r}$ reflect the spin currents entering the ferromagnet at each of the interfaces, providing the boundary conditions for the magnetic dynamics in the bulk. The electromotive forces in Eq.~\eqref{mf} (which enter in the modified Ohm's law) quantify the feedback applied by the magnet on the external electric circuit. Together these ingredients constitute self-consistent magnetoelectric dynamics, which we systematically address below.

{\em Nonlocal magnetoresistance}.|Our metallic contacts are integrated into an external electrical circuit such that in the series configuration [see Fig.~\ref{config}(a)] the two metals have currents running in the opposite directions, while in the parallel configuration [see Fig.~\ref{config}(b)] they run in the same direction. For a time-independent $\bE$, the hydrodynamic variables take, according to Eqs.~\eqref{EOMdamp}, a steady-state form: $\varphi(x,t)=f(x)+\W t$ and $n_z=\mbox{const}$~\cite{takeiPRL14}, where $\dot{\varphi}\equiv\W$ is the uniform global precession frequency of the magnetic texture [to be determined self-consistently from Eqs.~\eqref{EOMdamp} and the boundary conditions (see below)]. Matching the torques \eqref{torque} with the spin currents in the magnet near the two boundaries, we arrive at the following boundary conditions for $f(x)$:
\beq
\label{bc}
\begin{aligned}
-Af'(0)=\ez\cdot\bta'_l\ ,\ -Af'(L)=-\ez\cdot\bta'_r .
\end{aligned}
\eeq

For the series configuration, we have $\bj_l=-\bj_r=j\ey$. By inserting the steady-state form for $\varphi$ and $n_z$ into Eqs.~\eqref{EOMdamp} and \eqref{bc}, the rotation frequency within linear response becomes
\beq
\label{pf}
\W=\frac{\vartheta}{\g^{\up\down}+\g_\al/2}j\ ,
\eeq
with $\g_\al\equiv\al sL$. This result is analogous to that obtained in Ref.~\cite{takeiPRL14}, but now recast in terms of spin Hall phenomenology.
%Within linear-response, the boundary conditions then read
%\beq
%\label{bc}
%\begin{aligned}
%-Af'(0)&=\vartheta j-\g^{\up\down}\W\ ,\\
%-Af'(L)&=-\vartheta j+\g^{\up\down}\W\ .
%\end{aligned}
%\eeq
%The amount of spin current lost in the bulk due to Gilbert damping is given by $j^s(0)-j^s(L)=\al s\W L$~\cite{takeiPRL14}, which fixes the global frequency to 
%The adiabaticity condition (necessary for neglecting the $z$ component of the motive force) requires the time scale $2\p/\W$ to be much longer than the rc time of the circuit. Assuming the condition is satisfied, we find
In linear response, $\bve_{l,r}=\pm\vartheta\W\ey$, so that the modified Ohm's law gains an additional magnetoresistive contribution $(\rho+\rho_m) j=E$, where we obtain $|\bE_{l}|=|\bE_r|\equiv E$ under the assumption of identical metals and symmetrical interfaces, with 
\beq
\label{drho}
\rho_m=-\frac{\vartheta^2}{\g^{\up\down}+\g_\al/2}\ .
\eeq
The spin superfluidity thus reduces the effective resistivity of the circuit, implying that magnetic dynamics reduces net power dissipation, for a fixed current.

For the parallel configuration, $\bj_l=j_l\ey$ and $\bj_r=j_r\ey$, and the rotation frequency becomes, within linear response,
%\beq
%\begin{aligned}
%\label{bcdcp}
%-Af'(0)&=\vartheta j_l-\g^{\up\down}\W,\\
%-Af'(L)&=\vartheta j_r+\g^{\up\down}\W.
%\end{aligned}
%\eeq
\beq
\W=\frac{\vartheta}{\g^{\up\down}+\g_\al/2}\frac{j_l-j_r}{2}\ .
\eeq
Since the electrical circuit is parallel, $\bE_l=\bE_r$. Solving for the charge currents flowing in the metals, we obtain $j_l=j_r=E/\rho$, so that the magnetic texture is static, $\W=0$, a result consistent with the mirror symmetry about the $x=L/2$ plane. The resistivity of the electric circuit is not modified in this configuration. SSMR can be distinguished from the spin Hall magnetoresistance (SMR) recently discussed in the context of ferromagnet-metal interfaces~\cite{weilerPRL12,huangPRL12,nakayamaPRL13,vlietstraPRB13,chenPRB13,tserkovnyakPRB14}. SMR generates longitudinal corrections to electrical resistivity of order $\theta^2\la_N/t_N$ (when $t_N>\la_N$, the electron spin diffusion length) for both circuit configurations~\cite{chenPRB13,tserkovnyakPRB14}, contrasting with SSMR, which is nonzero only for the series configuration.

{\em Coherent ac spin transport}.|The ac regime can be explored by driving the metallic contacts by oscillating electric fields, i.e., $\bE_{l,r}(t)=\bE^{(0)}_{l,r}e^{-i\w t}$. In this section, we introduce the in-plane [U(1)-breaking] magnetic anisotropy by adding a free energy density term $\mathscr{F}_a=\ka\varphi^2/2$ (with $\kappa\ll K$), which augments Eq.~\eqref{EOMdamp} to 
\beq
\label{EOMdampan}
\dot\varphi=\frac{K}{s}n_z+\al\dot n_z\ ,\quad\dot n_z=\frac{A}{s}\varphi''-\al\dot\varphi-\frac{\ka}{s}\varphi\ .
\eeq
In the steady state, within linear response, the relevant hydrodynamic variables should oscillate at the ac frequency, such that $\varphi(x,t)=f(x)e^{-i \w t}$ and $n_z(x,t)=g(x)e^{-i \w t}$. The functions $f$ and $g$ can then be obtained using Eqs.~\eqref{bc} and \eqref{EOMdampan}. Throughout this section, we set $\bj_l(t)=j(t)\ey$ and $\bj_r=0$.
%, i.e., $\bE_l^{(0)}=E_0\ey$ and $\bE_r^{(0)}=0$.
%The electromotive force in the $z$ direction can again be neglected as long as the RC time is shorter than the ac time scale $2\p/\W$. Assuming this adiabaticity condition and 
Assuming a large eccentricity of precession, so that the ratio of torque coefficients is $|\vartheta/\eta|\gg |g/f|$, the electromotive force induced in the detector (right) contact reads $\bve_{r}(t)=i\vartheta\w f(L)e^{-i\w t}$. We then evaluate the transresistance as $\rho_t=\bve_r(t)/j(t)$.

\begin{figure}[t]
\centering
\includegraphics*[width=0.93\linewidth]{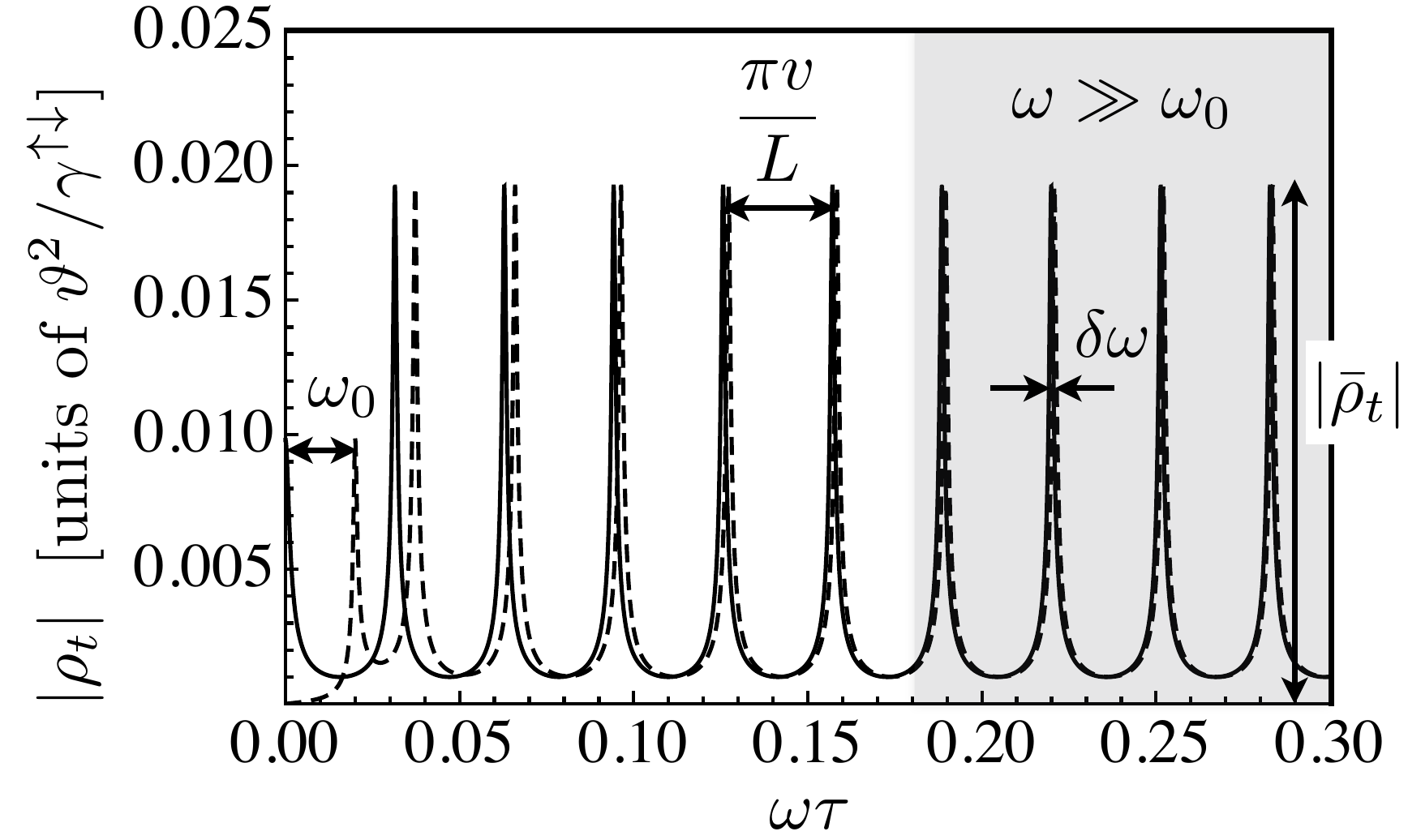}
\caption{Dynamic transresistance for $L=100\ell$ as a function of ac frequency $\w$. Here, we set $\al=10^{-3}$ and $\g^{\up\down}/s\ell=10^{-3}$. The solid line corresponds to the case without inplane anisotropy, while for the dashed line we use $\ka/K=4\times10^{-4}$.}
\label{fplot}
\end{figure}

There are two notable lengths scales, which determine the loss of spin transmission due to Gilbert damping. Previously, it was shown that a spin current carried by the zero-frequency mode (the superfluid component) decays algebraically as a function of system size $L$, and that the role of Gilbert damping becomes negligible for $L\ll L_\al\equiv\g^{\up\down}/\al s$~\cite{takeiPRL14}. For the easy-plane ferromagnet, spin current carried by a coherent finite-frequency spin wave should decay exponentially at distances larger than $\ell_\al=v\tau\al^{-1}\equiv\ell\al^{-1}$, where $v=\sqrt{AK}/s$ and $\tau^{-1}=K/s$ are the spin-wave velocity and easy-plane anisotropy, respectively. Since it is reasonable to assume $\g^{\up\down}/s\ell\ll1$, we have $L_\al\ll\ell_\al$. When $L\gg\ell_\al$, all spin-wave modes are strongly damped and the transresistance signal is exponentially small. We will therefore assume $L\ll\ell_\al$. In the ac regime, a series of resonance peaks appears in the (modulus of the) transresistance as a function of frequency (see Fig.~\ref{fplot}). For zero in-plane anisotropy (i.e., $\ka=0$) and a fixed $L$, these resonances occur at $\w=n\p v/L$, with integer $n$, such that the spin-wave velocity can be extracted from the peak intervals (see solid line in Fig.~\ref{fplot}). In the presence of in-plane anisotropy $\ka>0$ (see dashed line in Fig.~\ref{fplot}), the locations of the resonance peaks remain unperturbed for ac frequencies well above the gap scale, i.e., $\w\gg\w_0\equiv\sqrt{\ka K}/s$ (see the shaded region in Fig.~\ref{fplot}). Spin transmission is exponentially suppressed for $\w\ll\w_0$ and the first resonance peak is shifted to $\w_0$, allowing one to quantify the strength of the in-plane anisotropy. The resonant peaks have the square-root Lorentzian form, where the height $|\bar\rho_t|$ and the full-width half-maximum $\de\w$ (for $n\ge1$) are, respectively, given by 
\beq
\label{peakwidth}
|\bar\rho_t|=\frac{\vartheta^2}{2\g^{\up\down}+\g_\al/2}\ ,\quad\de\w=\frac{2\sqrt{3}}{s\tau L}\round{2\g^{\up\down}+\frac{\g_\al}{2}}\ .
\eeq
The ratio of peak heights between the $n\ge1$ resonances and the $n=0$ resonance is given by 
\beq
\label{ratio}
\frac{|\bar\rho_t|_{n\ge1}}{|\bar\rho_t|_{n=0}}=2\frac{\g^{\up\down}+\g_\al/2}{2\g^{\up\down}+\g_\al/2}\ .
\eeq
The overall factor of 2 and the factor of 2 in the denominator arise as a result of finite-frequency spin wave modes having antinodes at the two interfaces; this leads to an enhancement of interfacial damping (i.e., the factor of 2 in the denominator) and an overall increase in the transresistance signal (i.e., the overall factor of 2). The spin-mixing conductance, the spin Hall angle and the Gilbert damping parameter can all be experimentally determined for a given device via Eqs.~\eqref{peakwidth} and \eqref{ratio}. The measured values of $\de\w$ and the ratio determine the spin-mixing conductance and the Gilbert damping parameter. The spin Hall angle can finally be obtained from the measured peak height.

\begin{figure}[t]
\centering
\includegraphics*[width=0.94\linewidth]{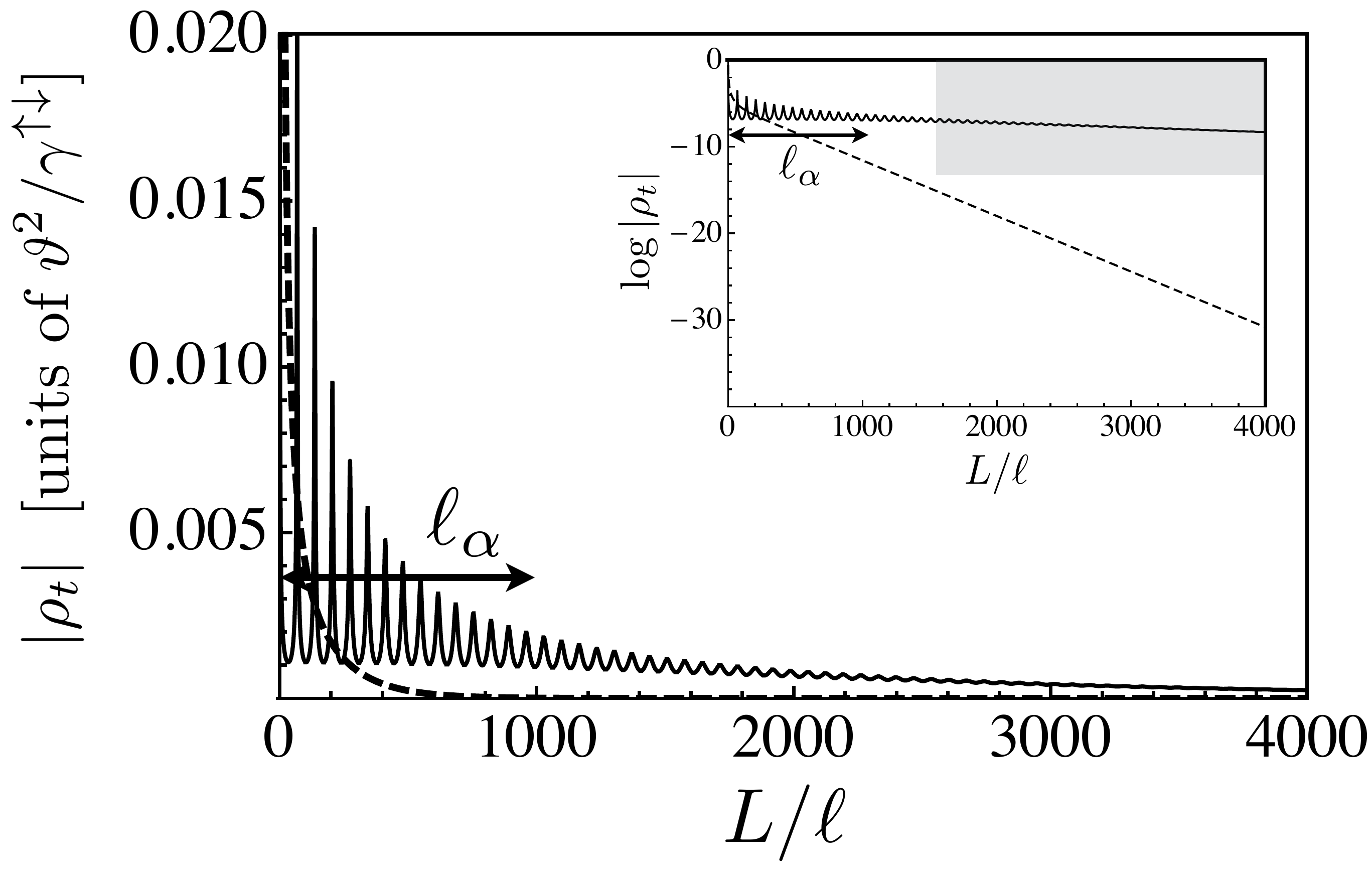}
\caption{Nonlocal magnetoresistance as a function of system size $L$. Here, we set $\al=10^{-3}$, $\g^{\up\down}/s\ell=10^{-3}$, and $\ka/K=4\times10^{-4}$ (corresponding to $\w_0\tau=0.02$). The solid line is for $\w\tau=0.05>\w_0\tau$, while the dashed line for $\w\tau=0.019<\w_0\tau$. The inset shows the semilog version of the same plot.}
\label{lplot}
\end{figure}

The dependence of transresistance $|\bar\rho_t|$ on magnet size $L$ reflects magnetic losses in the bulk. When the ac frequency exceeds the gap (see solid line in Fig.~\ref{lplot}), resonance peaks are observed up to the scale $L\sim\ell_\al$, but beyond this length scale, they become exponentially suppressed (as shown by the shaded region in Fig.~\ref{lplot} inset). For $L\ll\ell_\al$, the peak heights decay algebraically, as in the case of spin transmission via dc spin superfluidity [cf. Eq.~\eqref{drho}]. When $\w<\w_0$, spin-wave modes cannot be resonantly excited and the transresistance decays exponentially with increasing $L$ in a monotonic fashion (see dashed line in Fig.~\ref{lplot}). In the exponentially decaying region (when either $L\gg\ell_\al$ or $\omega<\omega_0$), the inverse decay length $\xi^{-1}$ (i.e., the slope of the semilog plot in the Fig.~\ref{lplot} inset) is given by
\beq
\xi^{-1}=\ell^{-1}\square{\De^2+\round{\al\w\tau}^2}^{1/4}\times
\begin{cases}
\cos\frac{\Theta}{2}\ ,& \w<\w_0\ ,\\
\sin\frac{\Theta}{2}\ ,& \w>\w_0\ ,
\end{cases}
\eeq
where $\De\equiv(\w^2-\w^2_0)\tau^2$ and $\Theta\equiv\tan^{-1}(\al\w\tau/|\De|)$.

{\em Conclusions.}|We studied the phase-coherent dc and ac spin transport through a magnetic insulator, sandwiched by two strongly spin-orbit coupled metals used for spin injection and detection. In the dc regime, we consider spin supercurrents carried by both dynamic and static magnetic textures, leading to distinct nonlocal magnetoresistances in the metals and a direct evidence for the spin superfluid state. We also address coherent spin transport in the ac regime, which allows one to extract static interfacial properties and the dynamic properties of the magnetic bulk, and to realize coherent spin transport even when the dc superfluid state is quenched by U(1)-breaking anisotropies.
%Studying dc superfluid spin transport in the context of the above two-terminal setup and in the presence of an $n$-fold anisotropy within the easy-plane~\cite{soninJETPL78} is an interesting question.

\begin{acknowledgments}
The authors would like to thank Amir Yacoby, Pramey Upadhyaya and Allan MacDonald for useful discussions. This work was supported by FAME (an SRC STARnet center sponsored by MARCO and DARPA) and in part by the Center for Emergent Materials, an NSF-funded MRSEC under Grant No. DMR-1420451, and the Kavli Institute for Theoretical Physics through Grant No. NSF PHY11-25915.
\end{acknowledgments}

\end{document}